\documentclass[a4paper,12pt]{article}
\usepackage{amsmath}
\usepackage{amssymb}
\font\black=cmbx10 \font\sblack=cmbx7 \font\ssblack=cmbx5
\font\blackital=cmmib10  \skewchar\blackital='177 \font\sblackital=cmmib7
\skewchar\sblackital='177 \font\ssblackital=cmmib5
\skewchar\ssblackital='177 \font\sanss=cmss10 \font\ssanss=cmss8 scaled
900 \font\sssanss=cmss8 scaled 600 \font\blackboard=msbm10
\font\sblackboard=msbm7 \font\ssblackboard=msbm5 \font\caligr=eusm10
\font\scaligr=eusm7 \font\sscaligr=eusm5 
\font\fraktur=eufm10 \font\sfraktur=eufm7 \font\ssfraktur=eufm5

\usepackage{color,soul}
\font\bsymb=cmsy10 scaled\magstep2
\def\all#1{\setbox0=\hbox{\lower1.5pt\hbox{\bsymb
       \char"38}}\setbox1=\hbox{$_{#1}$} \box0\lower2pt\box1\;}
\def\exi#1{\setbox0=\hbox{\lower1.5pt\hbox{\bsymb \char"39}}
       \setbox1=\hbox{$_{#1}$} \box0\lower2pt\box1\;}

\def\tx#1{{\fam0\relax#1}}

\newfam\bifam
\textfont\bifam=\blackital \scriptfont\bifam=\sblackital
\scriptscriptfont\bifam=\ssblackital

\newfam\blfam
\textfont\blfam=\black \scriptfont\blfam=\sblack
\scriptscriptfont\blfam=\ssblack

\newfam\bbfam
\textfont\bbfam=\blackboard \scriptfont\bbfam=\sblackboard
\scriptscriptfont\bbfam=\ssblackboard

\newfam\ssfam
\textfont\ssfam=\sanss \scriptfont\ssfam=\ssanss
\scriptscriptfont\ssfam=\sssanss
\def\ss#1{{\fam\ssfam\relax#1}}

\newfam\clfam
\textfont\clfam=\caligr \scriptfont\clfam=\scaligr
\scriptscriptfont\clfam=\sscaligr

\newfam\frfam
\textfont\frfam=\fraktur \scriptfont\frfam=\sfraktur
\scriptscriptfont\frfam=\ssfraktur

\font\frak=eufm10 scaled\magstep1
\def\goth#1{\hbox{{\frak#1}}}

\def\hpb#1{\setbox0=\hbox{${#1}$}
    \copy0 \kern-\wd0 \kern.2pt \box0}
\def\vpb#1{\setbox0=\hbox{${#1}$}
    \copy0 \kern-\wd0 \raise.08pt \box0}

\def\pmb#1{\setbox0\hbox{${#1}$} \copy0 \kern-\wd0 \kern.2pt \box0}
\def\pmbb#1{\setbox0\hbox{${#1}$} \copy0 \kern-\wd0
      \kern.2pt \copy0 \kern-\wd0 \kern.2pt \box0}
\def\pmbbb#1{\setbox0\hbox{${#1}$} \copy0 \kern-\wd0
      \kern.2pt \copy0 \kern-\wd0 \kern.2pt
    \copy0 \kern-\wd0 \kern.2pt \box0}
\def\pmxb#1{\setbox0\hbox{${#1}$} \copy0 \kern-\wd0
      \kern.2pt \copy0 \kern-\wd0 \kern.2pt
      \copy0 \kern-\wd0 \kern.2pt \copy0 \kern-\wd0 \kern.2pt \box0}
\def\pmxbb#1{\setbox0\hbox{${#1}$} \copy0 \kern-\wd0 \kern.2pt
      \copy0 \kern-\wd0 \kern.2pt
      \copy0 \kern-\wd0 \kern.2pt \copy0 \kern-\wd0 \kern.2pt
      \copy0 \kern-\wd0 \kern.2pt \box0}

\def\sT{{\ss T}}

\def\xd{\tx{d}}
\def\xi{\tx{i}}

\newcommand{\be}{\begin{equation}}
\newcommand{\ee}{\end{equation}}
\newcommand{\ra}{\rightarrow}

\newcommand{\bea}{\begin{eqnarray}}
\newcommand{\eea}{\end{eqnarray}}
\newcommand{\beas}{\begin{eqnarray*}}
\newcommand{\eeas}{\end{eqnarray*}}

\newcommand{\R}{\mathbb{R}}

\newcommand{\pa}{\partial}

\def\matriz#1#2{\left( \begin{array}{#1} #2 \end{array}\right) }

\mathchardef\za="710B  
\mathchardef\zb="710C  
\mathchardef\zg="710D  
\mathchardef\zd="710E  
\mathchardef\zve="710F 
\mathchardef\zz="7110  
\mathchardef\zh="7111  
\mathchardef\zvy="7112 
\mathchardef\zi="7113  
\mathchardef\zk="7114  
\mathchardef\zl="7115  
\mathchardef\zm="7116  
\mathchardef\zn="7117  
\mathchardef\zx="7118  
\mathchardef\zp="7119  
\mathchardef\zr="711A  
\mathchardef\zs="711B  
\mathchardef\zt="711C  
\mathchardef\zu="711D  
\mathchardef\zvf="711E 
\mathchardef\zq="711F  
\mathchardef\zc="7120  
\mathchardef\zw="7121  
\mathchardef\ze="7122  
\mathchardef\zy="7123  
\mathchardef\zf="7124  
\mathchardef\zvr="7125 
\mathchardef\zvs="7126 
\mathchardef\zf="7127  
\mathchardef\zG="7000  
\mathchardef\zD="7001  
\mathchardef\zY="7002  
\mathchardef\zL="7003  
\mathchardef\zX="7004  
\mathchardef\zP="7005  
\mathchardef\zS="7006  
\mathchardef\zU="7007  
\mathchardef\zF="7008  
\mathchardef\zW="700A  

\newcommand{\bepf}{\textit{Proof.-} }

\def\pd#1#2{\frac{\partial#1}{\partial#2}} 
\def\wt{\widetilde}
\def\li{{\textsf{\tiny L}}}
\def\ri{{\textsf{\tiny R}}}

\begin{document}

\title{Superposition rules, Lie theorem,\\ and partial differential equations}
\author{Jos\'e F. Cari\~nena\\
Depto. F\'{\i}sica Te\'orica, Universidad de Zaragoza\\
50009 Zaragoza, Spain\\
{\it e-mail:} jfc@unizar.es
\and
Janusz Grabowski\thanks{Research
supported by the Polish Ministry of Scientific Research and
Information Technology under the grant No. 2 P03A 036 25.}
\\
Mathematical Institute, Polish Academy of Sciences\\ ul. \'Sniadeckich
8, P. O. Box 21, 00-956 Warszawa, Poland\\ {\it e-mail:}
jagrab@impan.gov.pl
\and
Giuseppe Marmo\\
Dipartimento di Scienze Fisiche,
Universit\`a Federico II di Napoli\\
{\small and}\\
INFN, Sezione di Napoli\\
Complesso Universitario di Monte Sant'Angelo\\
Via Cintia, 80125 Napoli, Italy\\
{\it e-mail:} marmo@na.infn.it}
\date{}
\maketitle

\newtheorem{re}{Remark}
\newtheorem{theo}{Theorem}
\newtheorem{prop}{Proposition}
\newtheorem{lem}{Lemma}
\newtheorem{cor}{Corollary}
\newtheorem{ex}{Example}

\begin{abstract}
A rigorous geometric proof of the Lie's Theorem on nonlinear
superposition rules for solutions of non-autonomous ordinary
differential equations is given filling in all the gaps present in
the existing literature. The proof is based on an alternative but
equivalent definition of a superposition rule: it is considered as
a foliation with some suitable properties. The problem of
uniqueness of the superposition function is solved, the key point
being the codimension of the foliation constructed from the given
Lie algebra of vector fields. Finally, as a more convincing
argument supporting the use of this alternative definition of
superposition rule, it is shown that this definition allows an
immediate generalization of Lie's Theorem for the case of systems
of partial differential equations.




\end{abstract}

\section{Introduction}

The integration of systems of  differential equations  admitting
infinitesimal symmetries was the main concern of  Lie in
developing what is nowadays called the theory of Lie algebras and
Lie groups. In particular,  in a remarkable work \cite{LS} he was
able to prove an important  theorem connecting Lie algebras and
nonlinear superposition rules for solutions of some non-autonomous
systems of nonlinear ordinary differential equations. These
systems can be considered as a generalization of linear systems
but the superposition rule is no longer a linear function. Our aim
in this paper is to study once again, from a geometric viewpoint,
the theory of  systems of differential equations admitting a
(maybe nonlinear) superposition rule, allowing us to express the
general solution of the system by means of a superposition
function in terms of a (fundamental) set of  particular solutions,
with the hope of establishing clearly the necessary and sufficient
conditions for a system to admit such a superposition rule. This
time, however, we include also partial differential equations into our
considerations.

Even if the hypotheses of Lie's theorem were not accurately stated
from the today level of rigor, the resulting systems characterized
by means of an associated Lie algebra appear very often in
physical problems and in many cases the problem is related with
another one on the corresponding Lie group. This provides us with
both methods of reduction to simpler problems on one side, and
another method, introduced by Wei and Norman which involves some
algebraic manipulations based on Lie groups and Lie algebra
theories, on the other.

As far as we know, there is no rigorous proof of the if part in
Lie's theorem and the attempts known to us to get a rigorous
geometric proof share the same  pseudo-argument \cite{CGM00,{OS}}.
In this paper we prove that actually the existence of a
superposition rule for the solutions of a given non-autonomous
system implies that it has the explicit form which is usually
accepted. The proof is based on an alternative but equivalent
definition of superposition rule: we consider it as a foliation
with some appropriate properties explicitly formulated later on.
An auxiliary lemma is necessary to overcome the weak point in
previous derivations of the theorem \cite{CGM00,{OS}}.

On the other hand, the converse part is not given in its full
generality and almost nothing is said about the uniqueness of the
superposition function for these Lie systems. Only in \cite{NI} an
example given in \cite{LS}, for which there are two different
superposition functions, is pointed out. Our approach provides us
with an answer to this important question, the key point being the
codimension of the foliation constructed from the given Lie
algebra of vector fields. Moreover, this codimension is very
relevant when the action of the Lie algebra of vector fields on
the initial manifold is not transitive.

Finally, as a more convincing argument supporting the use of this
alternative definition of superposition rule, it  will be shown
that this definition allows an immediate generalization of Lie's
Theorem for the case of systems of partial differential equations.

The organization of the paper is as follows. Next section discuss
the concept of superposition function and gives a geometric
characterization of such superposition in terms of a foliation.
Section 3 is devoted to a complete proof of the statement of Lie's
theorem by establishing a lemma which allows us to overcome the
weak point of other previous derivations. The number of solutions in
a fundamental set is discussed in Section 4 and in Section 5 the
problem of uniqueness of the superposition rule is studied. Lie
systems on Lie groups and homogeneous spaces are considered in
Section 6 as most important examples. Finally, a generalization of
the Lie's Theorem for the case of systems of  first-order partial
differential equations is given in Section 7. An outlook with
future applications is given in the last section of the paper.

\section{Superposition rules for ordinary differential equations}

By a {\it superposition rule} (or a {\it superposition principle})
for a given system  of ordinary differential equations
\begin{equation}
\frac{dx^i}{dt }= Y^i(t,x)\,,\qquad  i=1,\ldots,n\,,\label{nonasys}
\end{equation}
one usually understands, after \cite{LS}, a superposition function
$\Phi:{\R}^{n(m+1)}\to {\R}^n$ given by
\be
x=\Phi(x_{(1)}, \ldots,x_{(m)};k_1,\ldots,k_n)\ ,\label{superpf}
\ee
such that the general solution can be written, at least for
sufficiently small $t$, as
\be
x(t)=\Phi(x_{(1)}(t), \ldots,x_{(m)}(t);k_1,\ldots,k_n)\ ,\label{superpft}
\ee
with $\{x_{(a)}(t)\mid a=1,\ldots,m\}$ being a fundamental set of
particular solutions of the system (\ref{nonasys}) and
$k=(k_1,\ldots,k_n)$ being a set of $n$  arbitrary constants
associated with  each particular solution.

The standard example is the system of linear
differential equations
\be
\frac {dx^i}{dt}=\sum_{j=1}^nA^i\ _j(t)\, x^j\ , \quad i=1,\ldots, n\,,
\label{lhsyst}
\ee
which admits the superposition function, with $m=n$,
$$x=\Phi(x_{(1)}, \ldots,x_{(n)};k_1,\ldots,k_n)=\sum_{i=1}^n k_i\, x_{(i)}\ .
$$
Of course, we can obtain every solution by superposing
$x_{(1)}(t),\ldots, x_{(m)}(t)$, for certain $k_1,\ldots,k_n$,
only if the functions $x_{(1)}(t),\ldots,x_{(m)}(t)$, are
appropriately independent, i.e. if they form a fundamental set of
solutions. In the above example it means that if the matrix
\begin{equation}
X(t)=(x^i_{(j)}(t))^i_j\qquad  i,j=1,\ldots,n\,,\label{ftalm}
\end{equation}
is invertible for small $t$.

The order in which the particular solutions are chosen is
irrelevant and therefore the superposition function should be such
that a permutation of two arguments only amounts to a change of
the parameters $k$. Note also that it is assumed that the
superposition function $\Phi$ does not depend explicitly on the
independent variable $t$, and this fact  has strong consequences
(see later the Lie theorem).

From a geometric perspective, systems of differential equations as
(\ref{lhsyst}) appear  as those determining the integral curves of
a $t$-dependent vector field in $\R^n$,
$$Y(t,x)=\sum_{i=1}^n Y^i(t,x)\,\pd{}{x^i}\,,$$
the generalization to the case of a $n$-dimensional manifold being
immediate. Note that then in any point $x\in N$, the $t$-dependent
vector field $Y$ in $N$, determines not only one vector but a
linear subspace, spanned by the set of vectors $\{Y(t,x)\mid t\in
\mathbb{R}\}$, of the corresponding tangent space. Actually, under
a time re-parametrization the vectors are rescaled and, when
changing the value of $t$, different vectors are obtained. In this
way it defines a `generalized' distribution for which the
dimension of the linear subspace can change from one point to
another. We will see later on that the case we are interested in
is such that the distribution defined by the $t$-dependent vector
field is involutive.

In order to look for superposition rules we need a  more geometric
picture. Let us first observe that, as a consequence of the
Implicit Function Theorem, the function $\Phi(x_{(1)},
\ldots,x_{(m)};\,\cdot\,):\R^n\to\R^n$ can be, at least locally
around generic points, inverted, so we can write
\begin{equation}
k=\Psi(x_{(0)}, \ldots,x_{(m)}) \label{defPsi}
\end{equation}
for a certain function $\Psi:{\R}^{n(m+1)}\to {\R}^n$.
Hereafter in order to handle a short notation we start writing $x_{(0)}$
instead of $x$. The foliation defined by the function $\Psi$ is now invariant
under permutations of the $(m+1)$ variables.

With some abuse of terminology we will also call the function $\Psi$ a
superposition function. The relation between $\Phi$ and $\Psi$ is given by:
\begin{equation}
k=\Psi(\Phi(x_{(1)}, \ldots,x_{(m)};k_1,\ldots,k_n),x_{(1)}, \ldots,x_{(m)})\,.\label{implth}
\end{equation}
For instance, for the system (\ref{lhsyst}) we have
$$\Psi(x_{(0)}(t), \ldots,x_{(n)}(t))=X^{-1}(t)x_{(0)}(t),$$
where $X(t)$ is the matrix given in (\ref{ftalm}). This example
indicates the obvious fact that, in general, the superposition
function $\Psi$ is defined on an open dense subset of
${\R}^{n(m+1)}$ rather than the whole ${\R}^{n(m+1)}$.

The fundamental property of the superposition function $\Psi$ is that as
\begin{equation}
k=\Psi(x_{(0)}(t),x_{(1)}(t), \ldots,x_{(m)}(t))\,,\label{kconstancy}
\end{equation}
the function  $\Psi(x_{(0)}, \ldots,x_{(m)})$ is constant on any
$(m+1)$-tuple of solutions of the system (\ref{nonasys}). This
property is true for any choice of $(m+1)$ solutions and this
means that the foliation is invariant under the permutation of the
$(m+1)$ arguments of the function $\Psi$.

After differentiation of relation (\ref{kconstancy}) with respect
to $t$, as the functions $x_i(t)$ are solutions of
(\ref{nonasys}), we get
\begin{equation}
D\Psi(Y(t,x_{(0)}), \ldots, Y(t,x_{(m)}))=0\,,\label{diffconst}
\end{equation}
i.e.
$$\sum_{i=1}^n\sum_{a=0}^m\pd\Psi{x^i_{(a)}}Y^i(t,x_{(a)})=0\,,
$$
and therefore the `diagonal prolongations' $\widetilde
Y(t,x_{(0)}, \ldots, x_{(m)})$ of the $t$-dependent  vector field
$Y(t,x)$, given by
$$\widetilde Y(t,x_{(0)}, \ldots,x_{(m)})=\sum_{a=0}^mY_a(t,x_{(a)})\,,\qquad
t\in {\R}\,,
$$
where
\begin{equation}\label{Ya}
Y_{a}(t,x_{(a)})=\sum_{i=1}^n Y^i(t,x_{(a)})\,\pd{}{x^i_{(a)}}
\end{equation}
are $t$-dependent  vector fields on ${\R}^n\times \cdots\times {\R}^n$ ($(m+1)$ factors)
which are tangent to the level sets of $\Psi$ as displayed by (\ref{diffconst}).

The level sets of $\Psi$
corresponding to
regular values define a $n$-codimensional foliation $\mathcal{F}$ on an open
dense subset $U\subset {\R}^n\times \cdots\times {\R}^n$ ($(m+1)$ factors)
and the family $\{\widetilde Y(t),\,t\in {\R}\}$ of vector
fields in ${\R}^{n(m+1)}$   consists of vector fields tangent to
the leaves of this foliation.

This foliation has another important property. Since on the level
set $\mathcal{F}_k$ corresponding to $k=(k_1,\ldots,k_n)\in{\R}^n$
and given $(x_{(1)},\ldots,x_{(m)}) \in {\R}^{nm}$, there is a
unique point $(x_{(0)},x_{(1)},\ldots,x_{(m)})\in \mathcal{F}_k$,
namely,
$(\Phi(x_{(1)},\ldots,x_{(m)};k),x_{(1)},\ldots,x_{(m)})\in
\mathcal{F}_k $ (cf. (\ref{implth})), then the projection onto the
last $m$ factors
$${\rm pr}:(x_{(0)},x_{(1)},\ldots,x_{(m)})\in {\R}^{n(m+1)}\mapsto
(x_{(1)},\ldots,x_{(m)})\in {\R}^{nm}
$$
induces diffeomorphisms on the leaves $\mathcal{F}_k$ of $\mathcal{F}$.

This can also be viewed as the fact that the foliation
$\mathcal{F}$ corresponds to a connection $\Delta$ in the bundle
${\rm pr}: {\R}^n\times \cdots\times {\R}^n={\R }^{n(m+1)}\to
{\R}^n\times \cdots\times {\R}^n={\R }^{nm}$ with trivial
curvature. The restriction of the projection ${\rm pr}$ to a leaf
gives a one-to-one map. In this way there is a linear map among
vector fields in ${\R }^{nm}$ and (horizontal) vector fields
tangent to a leaf.

Note that the knowledge of this connection (foliation) gives us
the superposition principle without referring to the function
$\Psi$ (which we can change by composing it, for instance, with a
diffeomorphism of ${\R}^n$): if we fix the point $x_{(0)}(0)$,
i.e. we choose a $k=(k_1,\ldots,k_n)$, and $m$ solutions
$x_{(1)}(t),\ldots,x_{(m)}(t)$, then $x_{(0)}(t)$ is the unique
point in ${\R}^n$ such that
$(x_{(0)}(t),x_{(1)}(t),\ldots,x_{(m)}(t))$ belongs to the same
leaf of $\mathcal{F}$ as
$(x_{(0)}(0),x_{(1)}(0),\ldots,x_{(m)}(0))$. Thus, it is only
$\mathcal{F }$ that really matters when the superposition rule is
concerned.

On the other hand, if we have a connection $\nabla$ in the bundle ${\rm pr}:{\R}^{n(m+1)}
\to {\R}^{nm}$ with a trivial curvature, i.e. we have a horizontal distribution
$\nabla$ in $ T{\R}^{n(m+1)}$ that is involutive, which therefore
can be integrated
to give a foliation in ${\R}^{n(m+1)}$ such that the vector fields $\widetilde
Y(t)$ belong to $\nabla$ (equivalently, are tangent to $\mathcal{F}$, i.e. are
horizontal), then the procedure described above determines a superposition rule
for the system (\ref{nonasys}).

Indeed, let $k\in {\R}^n$ enumerate smoothly the leaves
$\mathcal{F}_k$ of $\mathcal{F}$ (e.g. by  a choice of a small
cross-section of $\mathcal{F}$), then
$\Phi(x_{(1)}(t),\ldots,x_{(m)}(t);k)$ defined as the unique point
$x_{(0)}(t)\in{\R}^n$ such that
$(x_{(0)}(t),x_{(1)}(t),\ldots,x_{(m)}(t))\in \mathcal{F}_k$, is a
superposition rule for the system (\ref{nonasys}) of ordinary
differential equations. To see this, let us observe that the
inverse is $\Psi(x_{(0)}(t),x_{(1)}(t),\ldots,x_{(m)}(t))=k$,
which is equivalent to
$(x_{(0)}(t),x_{(1)}(t),\ldots,x_{(m)}(t))\in \mathcal{F}_k$. If
we fix $k$ and take solutions $x_{(1)}(t),\ldots,x_{(m)}(t)$ of
(\ref{nonasys}), then $x_{(0)}(t)$ defined by the condition
$\Psi(x_{(0)}(t),x_{(1)}(t),\ldots,x_{(m)}(t))=k$ satisfies
(\ref{nonasys}). Indeed, let $x'_{(0)}(t)$ be the solution of
(\ref{nonasys}) with initial value $x'_{(0)}(0)= x_{(0)}(0)$.
Since the $t$-dependent vector fields $\widetilde Y(t)$ are
tangent to $\mathcal{F}$, the curve
$(x'_{(0)}(t),x_{(1)}(t),\ldots,x_{(m)}(t))$ lies entirely on a
leaf of $\mathcal{F}$, so on $\mathcal{F}_k$. But the point of one
leaf is entirely determined by its projection ${\rm pr}$, so
$x'_{(0)}(t)=x_{(0)}(t)$ and $x_{(0)}(t)$ is a solution. Thus we
have proved the following geometric characterization of
superposition rules:

\begin{prop}
Giving a superposition rule (\ref{superpf})
for a system of differential equations
(\ref{nonasys}) is equivalent to giving a zero curvature connection in the
bundle ${\rm pr}:{\R}^{(m+1)n}\to {\R}^{nm}$ for which the diagonal
prolongations $\widetilde Y(t)$ of the $t$-dependent vector fields $Y(t),\, t\in {\R}$,
defining the system (\ref{nonasys}) are horizontal.
\end{prop}
Note that we can also consider arbitrary manifolds $N$ instead of
${\R}^n$. The superposition functions are then given by maps
$\Phi:N^{m+1}\to N$ or by  appropriate foliations in $N^{m+1}$,
i.e. zero curvature connections in the bundle  ${\rm
pr}:N^{m+1}\to N^m$.

\begin{re}{\rm
Frankly speaking, the connection is defined only generically,
usually over an open-dense subset. But this is a general problem
with superpositions, which hold only `generically'.} In the sequel
all objects and constructions will be `generic' in this sense.
\end{re}

\begin{ex}{\rm
Consider the (generalized) foliation $\mathcal{F}$ of codimension
one generated by the vector field $x\, \partial /\partial x+y\,
\partial /\partial y$ on ${\R}^2-\{(0,0)\}$. It defines a zero curvature connection for the
bundle ${\rm pr}: {\R}\times ({\R}-\{0\})\to {\R}-\{0\}$, ${\rm
pr}(x,y)=y\in {\R}$. The leaves of $\mathcal{F}$ are of the form
$(e^{t}x,e^ty), \, t\in {\R}, \, y\ne 0$. In particular, the
function $\Psi(x,y)=x/y$ is constant on the leaves. The diagonal
prolongation of the $t$-dependent vector fields $a(t)\, x\,
\partial /\partial x$ are of the form $a(t)\,(x\,
\partial /\partial x+y\, \partial /\partial y)$ and are tangent to
$\mathcal{F}$. This gives us  the superposition rule for the linear differential
equation
$\dot x=a(t)\, x$
as follows: If $x_{(1)}(t)$ is a solution, then $(x_{(0)}(t),x_{(1)}(t))$
belongs to one leaf, e.g with $x/y=k$, if and only if $x_{(0)}(t)=k\,
x_{(1)}(t)$. This is the standard superposition rule for this equation:
$$\Phi(x_{(1)}(t);k)=k\, x_{(1)}(t)\,.
$$}
\end{ex}

\begin{re}{\rm
It is clear from the above proposition that when the diagonal
prolongations $\widetilde Y(t)$ generate a foliation of dimension
smaller than the dimension of $\mathcal{F}$, we can change
$\mathcal{F}$ (i.e. we can change the connection $\nabla$)
respecting the required properties. This means that the
corresponding system (\ref{nonasys}) admits many different
superposition rules even if we regard compositions of $\Phi$ with
diffeomorphisms of ${\R}^n$ as equivalence relations.}
\end{re}

\section{Lie theorem for ODE's systems admitting superposition rules}

In this section we shall give a proof of the classical Lie's
theorem on ODE's admitting a superposition rule (cf.
\cite{{LS},{CGM00},{CGM01}, {CarRamGra}}).

First of all, in this proof we shall  fill  in all gaps which
appear in the known literature. Second, we shall use the
alternative characterization of superposition rules given in
Proposition 1 and last, but not least, the proof, as it will be
presented, can be immediately extended to the case of partial
differential equations.

Remark first that all considerations will be `generic' and local.
Let us recall that  a foliation $\mathcal{F}$ of an open dense
subset of $\widetilde N=N^{m+1}=N\times \cdots\times N$ ($(m+1)$
factors) defines a superposition rule for the system of
differential equations (\ref{nonasys}), if and only if
$\mathcal{F}$ is of codimension $n$, the projection ${\rm
pr}:\widetilde N\to N\times \cdots\times  N=N^m$ (only $m$
factors) onto the last $m$ arguments maps leaves of  $\mathcal{F}$
diffeomorphically and, furthermore, the generalized foliation
$\mathcal{F}_0$ generated by the family $\{\widetilde Y(t)\mid
t\in{\R}\}$ of diagonal prolongations of $Y(t)$ is contained in
$\mathcal{F}$.

We shall work only with the regular part of $\mathcal{F}_0$. Such
regular part is spanned by $\{\widetilde Y(t)\mid t\in{\R}\}$,
i.e. in any case, it is spanned by diagonal prolongations of some
vector fields on $N$  as $[\widetilde Y(t),\widetilde
Y(t')]=\widetilde{[Y(t), Y(t')]}$, etc. Let $\widetilde X_1,
\ldots, \widetilde X_r$ be diagonal prolongations spanning locally
the regular part of $\mathcal{F}_0$ of dimension $r$, and
therefore ${\rm pr}_*(\widetilde X_1),\ldots, {\rm
pr}_*(\widetilde X_r)$ are assumed to be linearly independent at a
generic point. We clearly have $r\leq mn$.

Since $\widetilde X_1, \ldots, \widetilde X_r$ span an
$r$-dimensional foliation, then
$$[\widetilde X_\alpha,\widetilde X_\beta]=\sum_{\gamma=1}^rc_{\alpha\beta}\,^\gamma\
\widetilde X_\gamma\,,
$$
for some $r^3$ functions $c_{\alpha\beta}\,^\gamma$ defined on
$\widetilde N$. Note also that $[\widetilde X_\alpha,\widetilde
X_\beta]$ are diagonal prolongations as brackets of  diagonal
prolongations and the projections ${\rm pr}_*(\widetilde
X_1),\ldots, {\rm pr}_*(\widetilde X_r)$, are assumed to be
functionally independent. Then we shall use the following lemma.

\begin{lem}
Let $\widetilde X_\alpha={\displaystyle
\sum_{a=0}^m}X_{\alpha(a)}$, for $\alpha=1,\ldots r$, and with
$r\leq mn$, be diagonal prolongations  to $\widetilde N$ of vector
fields $X_\alpha$ on $N$,  and such that at each point $p$ of
$N^m$ the vectors that are the projections of their values, ${\rm
pr}_*(\widetilde X_\alpha)(p)
={\displaystyle\sum_{a=1}^m}X_{\alpha(a)}(p)$, are linearly
independent. Then, ${\displaystyle \sum_{\alpha=1}^r}b_\alpha \,
\widetilde X_\alpha$, with $b_\alpha\in C^\infty (\widetilde N)$,
is again a diagonal prolongation if and only if the coefficients
$b_\alpha$ are constant.
\end{lem}

\bepf Let us write in local coordinates
$$X_\alpha=\sum_{i=1}^n A_\alpha^i(x)\pd{}{x^i}\,,
$$
which implies that
$$
\widetilde X_\alpha=\sum_{i=1}^n \sum_{a=0}^m A_\alpha^i(x_{(a)})\pd{}{x^i_{(a)}}\,.
$$
Then,
$$\sum_{\alpha=1}^r b_\alpha (x_{(0)},\ldots, x_{(m)})\widetilde
X_\alpha=\sum_{\alpha=1}^r\sum_{i=1}^n\sum_{a=0}^m b_\alpha
(x_{(0)},\ldots, x_{(m)}) A_\alpha^i(x_{(a)})\,\pd{}{x_{(a)}^i}\,,
$$
which is a diagonal prolongation if and only if there are
functions $B_a^i(x)$, for $a=0, \ldots, m$, and $i=1,\ldots,n$,
such that for each pair of indexes $i$ and $a$,
$$\sum_{\alpha=1}^rb_\alpha (x_{(0)},\ldots,
x_{(m)})\,A_\alpha^i(x_{(a)})= B_a^i(x_{(a)})\ ,\quad
a=0,\ldots,m\,\ i=1,\dots,n.
$$
In particular, the $r$ functions $b_\alpha (x_{(0)},\ldots,
x_{(m)})$ solve the following subsystem of linear equations in the
unknown $u_\alpha$, for $\alpha=1,\ldots,r$:
\begin{equation}
\sum_{\alpha=1}^r u_\alpha\,A_\alpha^i(x_{(a)})=
B_a^i(x_{(a)})\,,\qquad {\rm with \ } i=1,\ldots,n\,,\
a=1,\ldots,m\,.\label{linsyst}
\end{equation}
But the rank of the matrix $(A_\alpha^i(x_{(a)}))_\za^{i,a}$ is
$r\leq mn$ as the projections ${\rm pr}_*(\widetilde X_1), \ldots,
{\rm pr}_*(\widetilde X_r)$ are assumed to be linearly
independent. Thus the solutions $u_1,\ldots,u_r$ of
(\ref{linsyst}) are unique and are completely determined by the
matrix $A_\alpha^i(x_{(a)})_\za^{i,a}$ and the vector
$B_a^i(x_{(a)})^{i,a}$, with $a=1,\ldots,m$, so they do not depend
on $x_{(0)}$.  But since the diagonal prolongations are invariant
with respect to the symmetry group $S_{m+1}$ acting on $\widetilde
N= N^{m+1}$ in an obvious way, the functions $b_\alpha
(x_{(0)},\ldots, x_{(m)})$ do not depend also on the other
variables $x_{(1)},\ldots, x_{(m)}$.

\begin{re}{\rm Let us note that the assumption on the projections is crucial
for the above lemma and actually without such assumption the
result of this fundamental lemma is simply wrong as the following
example shows. Consider the following two vector fields in
${\R}^2$ which are prolongations of vector fields in ${\R}$:
$$
\widetilde X_1= \pd{}x+\pd{}y\,,\qquad \widetilde X_2= x\pd{}x+y\pd{}y\,,
$$
and the functions
$$b_1(x,y)=x\,y\,,\qquad b_2(x,y)=-(x+y)$$
for which
$$
b_1(x,y)\widetilde X_1(x,y)+b_2(x,y)\,\widetilde X_2(x,y)= -\left(x^2\pd{}x+y^2\pd{}y\right)
$$
is also a prolongation. However the coefficients $b_1$ and $b_2$
are not constant. This is the standard gap in the proofs of Lie's
theorem we found in the literature. One usually claims that a
functional combination of diagonal prolongations is a diagonal
prolongation only if the coefficients are constant without
assuming that the corresponding projections are linearly
independent.}
\end{re}
Now, we can now prove the above mentioned Lie theorem using the
previous results.

\begin{theo}  The system (\ref{nonasys}) on a differentiable manifold $N$
admits a superposition rule if and only if the $t$-dependent vector field
$Y(t,x)$ can be locally written in the form
$$Y(t,x)=\sum_{\alpha=1}^r b_\alpha(t)\, X_\alpha(x)$$
where the $t$-dependent vector fields $X_\alpha$, $\za=1,\dots,
r$, close on a finite-dimensional real Lie algebra, i.e. there
exist $r^3$ real numbers $c_{\alpha\beta}\,^\gamma $ such that
$$[X_\alpha,X_\beta]= \sum_{\gamma=1}^r c_{\alpha\beta}\,^\gamma\,
X_\gamma\,,\qquad \forall \alpha,\beta=1,\ldots,r\,.
$$
\end{theo}

\bepf  Suppose that the system admits a superposition rule and let $\mathcal{F}$ be
the foliation corresponding to the superposition function. We know
already that the generators $\{\widetilde X_\alpha\mid
\alpha=1,\ldots, r\}$ of the regular part of
$\mathcal{F}_0\subset\mathcal{F}$ close on a Lie algebra
\begin{equation}
[\widetilde X_\alpha,\widetilde X_\beta]=\sum_{\gamma=1}^r c_{\alpha\beta}\,^\gamma \widetilde
X_{\gamma},
\label{invol}
\end{equation}
where the coefficients $c_{\alpha\beta}\,^\gamma $ are constant, so also
$$[X_\alpha, X_\beta]=\sum_{\gamma=1}^r c_{\alpha\beta}\,^\gamma\,
X_{\gamma}\,.
$$
Since every $ \widetilde Y(t)$ is tangent to $\mathcal{F}_0$ there are functions
$b^\alpha_t(x_{(0)},\ldots, x_{(m)})$ such that
$$\widetilde Y(t)=\sum_{\alpha=1}^rb_t^\alpha \,\widetilde X_\alpha\,.
$$
But $\widetilde Y(t)$ is a diagonal prolongation, so, using the
fundamental lemma once more, we get that the
$b_t^\alpha=b^\alpha(t)$ are independent on $x_{(0)},\ldots,
x_{(m)}$. Hence
\begin{equation}
\widetilde Y(t)=\sum_{\alpha=1}^rb^\alpha(t)\, \widetilde X_\alpha \label{wty}
\end{equation}
and also
\begin{equation}
Y(t)=\sum_{\alpha=1}^rb^\alpha(t)\, X_\alpha\,.\label{ydet}
\end{equation}

To prove the converse property,  assume that the $t$-dependent
vector field $Y(t,x)$ can be written as in (\ref{wty}) and  define
$\wt Y(t)$ by (\ref{ydet}). We can additionally assume that the
vector fields $ X_\alpha$ are linearly independent over $\R$. Thus
they define an $r$-dimensional Lie algebra with structure
constants $c_{\alpha\beta}\,^\gamma$ (and the corresponding simply
connected Lie group if the vector fields are complete).

Since only non-trivial functional dependence of $X_1,\ldots,X_r$
is possible, there is a number $m\leq r$ such that their diagonal
prolongations to $N^m=N\times \cdots\times N$ ($m$ times) are
generically linearly independent at each point. The distribution
spanned by the diagonal prolongations $\widetilde X_1,
\ldots,\widetilde X_r$ to $\widetilde N=N^{m+1}$ is clearly
involutive, so it defines an $r$-dimensional foliation
$\mathcal{F}_0$ of $\widetilde N$. Moreover, the leaves of this
foliation  project onto the product of the last $m$ factors
diffeomorphically and they are at least $n$-codimensional. Now, it
is obvious that we can extend this foliation to an
$n$-codimensional foliation $\mathcal{F}$ with the latter
property, and this foliation, according to proposition 1, defines
a superposition rule. Here, of course, extending of a foliation
means that the leaves of the smaller are submanifolds of the
leaves  of the extension.

\begin{re}{\rm There is another way to look at the superposition the way  we
proposed.  We can consider two projections: the first one, ${\rm
pr}:\widetilde N=N^{n(m+1)}\to N^{nm}$ is on the last $m$ factors
and the second one,  ${\rm pr}_1: \widetilde N\to N$ the
projection on the first factor. These projections are clearly
transversal and the first one is a fibration if the foliation
$\mathcal{F}$ is conserved, i.e. every curve in $N^m$ has a unique
lifting in a fixed leaf of $\mathcal{F}$. Then, the superposition
associated with this lift is just ${\rm pr}_1$ of the lift of
particular solutions.}
\end{re}
\begin{re}{\rm
We hope that it is clear to the reader that in our picture of
superposition rules there is no real need to take all the
manifolds $N$ equal in the product where the superposition
foliation lives. We can consider as well $\widetilde
N=N_0\times\cdots N_m$ with analogous projection and fibration
property. This means that we can get a solution of a system on
$N_0$ out of solutions of some systems on $N_a$, $a=1,\ldots,m$.
This is a geometric picture for the Darboux (B\"acklund)
transformations. We will, however, discuss this problem in a
separate paper.}
\end{re}

\section{Determination of the number $m$ of solutions of a fundamental set}

Our proof of the Lie's Theorem contains an information about the
number $m$ of solutions involved in the superposition rule. For a
Lie system defined by a $t$-dependent vector field of the form
$$
Y(t,x)=\sum_{\alpha=1}^r b_\alpha(t)\, X_\alpha(x)\,,
$$
with generic $b_\za(t)$ the number $m$ turned out to be the
minimal $k$ such that the diagonal prolongations of
$X_1,\dots,X_r$ to $N^k$ are linearly independent at (generically)
each point: the only real numbers solution of the linear system
$$\sum_{\alpha=1}^r c_\alpha\, X_\alpha(x_{(a)})=0\,,\qquad a=1,\ldots ,k
$$
at a generic point $(x_{(1)},\dots,x_{(k)})$ is the trivial
solution $c_\alpha=0$, $\alpha=1,\ldots,m$,  for $k=m$ and there
are nontrivial solutions for $k<m$.

For instance, for the Riccati equation
$$ \dot x=b_0(t)+b_1(t)\, x+b_2(t)x^2\,,
$$ in which the vector fields generating the foliation $\mathcal{F}_0$
are prolongations of the vector fields
\be\label{ric}X_0=\frac{\pa}{\pa x}\,,\qquad X_1=x\frac{\pa}{\pa
x}\,,\qquad X_2=x^2\frac{\pa}{\pa x}\,,
\ee
representing an $sl(2,\R)$-action, we see that the system
$$c_0+c_1x_1+c_2x_1^2=0\,,\qquad c_0+c_1x_2+c_2x_2^2=0
$$
has a nontrivial solution but the one given by
$$c_0+c_1x_1+c_2x_1^2=0\,,\qquad c_0+c_1x_2+c_2x_2^2=0\,,\qquad
c_0+c_1x_3+c_2x_3^2=0\,,
$$
does not admit non-trivial solutions because the determinant of
the coefficients is invertible when the three points $x_1$, $x_2$
and $x_3$ are different. This implies that $m=3$ in the
superposition rule for the Riccati equation.

\section{Nonuniqueness of the superposition rule}

In some cases the foliation $\mathcal{F}_0$ spanned by the
prolongations of $t$-dependent vector fields defining the dynamics
is already $n$-codimensional and we get a unique (minimal)
superposition rule (e.g. Riccati equation). In the cases with
codim\,$\mathcal{F}_0>n$, we have some ambiguity in choosing the
superposition rule as we can extend $\mathcal{F}_0$ to an
$n$-codimensional foliation in different ways.

\begin{ex}{\rm
Consider the action of the Abelian group $\R$ acting on $N=\R^2$
by horizontal translations., i.e. $X_1=\partial/\partial x$. This
action on ${\R}^2$  is free and we have $m=1$, so $\widetilde
N={\R}^2\times {\R}^2$ and $\mathcal{F}_0$ is spanned by
$\partial/\partial x_{(0)}+\partial/\partial x_{(1)}$, where the
coordinates in $\widetilde N$ are denoted $( x_{(0)}, y_{(0)},
x_{(1)}, y_{(1)})$. We can extend such foliation to a
2-dimensional foliation $\mathcal{F}$ of $\widetilde N$ with the
required property with respect to the projection ${\rm pr}(
x_{(0)}, y_{(0)}, x_{(1)}, y_{(1)})=( x_{(1)}, y_{(1)})$ in
different ways. For instance, we can take $\mathcal{F}$ to be
given by the level sets of the mapping
$$F( x_{(0)}, y_{(0)}, x_{(1)}, y_{(1)})=( x_{(0)}-x_{(1)},f( y_{(0)},
y_{(1)}))$$ with $f$ being an arbitrary function such that
$\partial f/\partial y_{(0)}\ne 0$. Then, every solution $(
x_{(1)}(t), y_{(1)}(t)=y_{(1)}(0))$ of the system of differential
equations
$$\dot x=a(t)\,,\qquad \dot y=0\,,$$
gives a new solution $( x_{(0)}(t), y_{(0)}(t)=y_{(0)}(0))$
associated with the level set  of $(k_1,k_2)$ by
$$( x_{(0)}(t)= x_{(1)}(t)+k_1,y_{(0)}(t)=y_{(0)}(0))\,.$$
where $y_{(0)}(0)$ is the unique point in $\R$ satisfying
$$f( y_{(0)}(0),y_{(1)}(0))=k_2\,.
$$
In the case $f( y_{(0)},y_{(1)})=y_{(0)}-y_{(1)}$  we recover the
`standard' superposition rule:
$$
\Phi(x_{(1)}, y_{(1)};k_1,k_2)=(x_{(1)}+k_1, y_{(1)}+k_2)\,.
$$}
\end{ex}

\begin{ex}{\rm A very simple example is given by the separable first-order
differential equation
$$\dot x=a(t)\, f(x)\,,$$
with $a$ and $f$ being arbitrary smooth functions, and where $f$
is assumed to be of a constant sign (otherwise we can restrict
ourselves to a neighbourhood of a point in which $f$ does not
vanish). In this case $N=\mathbb{R}$ and we can consider the
vector field in $\mathbb{R}$
$$X(x)=f(x)\,\pd{}x\,.
$$
As the function $f$ does not vanish, we have  $m=1$ and the
diagonal prolongation
$$\widetilde X(x_{(0)},x_{(1)})=f(x_{(0)})\,\pd{}{x_{(0)}}+
f(x_{(1)})\,\pd{}{x_{(1)}}
$$
generates a one-dimensional foliation in $\mathbb{R}^2$ whose
leaves are the level sets of a function $\Psi(x_{(0)},x_{(1)})$
such that
$$
f(x_{(0)})\,\pd{\Psi}{x_{(0)}}+
f(x_{(1)})\,\pd{\Psi}{x_{(1)}}=0\,,
$$
which gives rise to the following characteristic system
$$\frac{dx_{(0)}}{f(x_{(0)})}=\frac{dx_{(1)}}{f(x_{(1)})}\,.
$$
Therefore, if the function $\phi(y)$ is defined by
$$\phi(y)=\int^y_0\frac{d\zeta}{f(\zeta)}\,,
$$
then we find that the leaves are characterized by a constant $k$
in such a way that
$$\phi(x_{(0)})-\phi(x_{(1)})=k\,.
$$
The function $\phi$ is a monotone function, because
$\phi'(x)=f(x)$ and $f(x)$ has constant sign. Therefore, there
exists an inverse function which allows to write the superposition
rule as
$$x=\phi^{-1}\left(k+\phi(x_{(1)})\right)\,.
$$
For instance, if $f(x)=1/x^2$, we find that $\phi(x)=-1/x
=\phi^{-1}(x)$, an we obtain the following superposition rule.
$$x=\frac{x_{(1)}}{1-k\,x_{(1)}}\,.
$$}
\end{ex}

\begin{ex}{\rm
It has been pointed out in \cite{NI2} the following example of the
original Lie's work:
$$\left\{
\begin{array}{rcl}{\displaystyle\frac{dx}{dt}}&=&a_{12}(t)\,
y+b_1(t)\\
{\displaystyle\frac{dy}{dt}}&=& -a_{12}(t)\,x+b_2(t)
\end{array}\right.
$$
In principle, it is a particular example of an inhomogeneous
linear system and we expect to have an affine superposition rule
involving three different solutions:
$$x=\Phi_1(x_{(1)},x_{(2)},x_{(3)})=x_{(1)}+k_1(x_{(2)}-x_{(1)})+k_2(x_{(3)}-x_{(1)})\,.
$$
However, we can obtain a superposition rule which is not linear but involves
only two solutions. It is due to the fact that here is not the affine group in
two dimensions which is play the relevant r\^ole, but the Euclidean group.
The foliation corresponding to this Lie system is generated by the prolongations
of the vector fields
$$X_1=\pd{}{x}\,,\qquad X_2=\pd{}{y}\,,\qquad X_3=y\pd{}x-x\pd{}y\,.
$$
First of all, $m$ is different from 1, because there exist
nontrivial coefficients $\lambda_1$, $\lambda_2$ and $\lambda_3$
such that $\lambda_1\, X_{1}(x_{(1)})+ \lambda_2\,
X_{2}(x_{(1)})+\lambda_3\, X_{3}(x_{(1)})=0$, at a given point
$x_{(1)}$, for instance, $\lambda_1=-y_{(1)}$,
$\lambda_1=x_{(1)}$, $\lambda_3=1$. However, the only coefficients
$\lambda_1$, $\lambda_2$ and $\lambda_3$ such that
$$\lambda_1\, X_1(x_{(1)})+\lambda_2\, X_2(x_{(1)})+\lambda_3\,
X_3(x_{(1)})=0\,,\qquad \lambda_1\, X_1(x_{(2)})+\lambda_2\, X_2(x_{(2)})+\lambda_3\,
X_3(x_{(2)})=0\,,
$$
with $x_{(1)}\ne x_{(2)} $ are $\lambda_1=\lambda_2=\lambda_3=0$ and therefore $m=2$.

The function $\Psi$ defining the superposition rule satisfies
$\wt X_1\Psi=\wt X_2\Psi=\wt X_3\Psi=0$
with
$$\wt X_1=\pd{}{x}+\pd{}{x_{1}}+\pd{}{x_{2}}\,,\qquad \wt
X_2=\pd{}{y}+\pd{}{y_{1}}+\pd{}{y_{2}}\,,
$$
and
$$ \wt  X_3=y\pd{}x-x\pd{}y+y_1\pd{}{x_1}-x_1\pd{}{y_1}+y_2\pd{}{x_2}-x_2\pd{}{y2}\,.
$$
The two first conditions imply that $\Psi$ must be of the form
$$\Psi(x_{0},y_{0},
x_{1},y_{1},x_{2},y_{2})=\psi(x_{0}-x_{1},x_{0}-x_{2},
y_{0}-y_{1},y_{0}-y_{2})\,,
$$
what suggests the change of variables
$$u_1=x_0-x_1,\quad u_2=x_0-x_2,\quad u_3=x_0,\quad v_1=y_0-y_1,\quad
v_2=y_0-y_2,\quad v_3=y_0\,,
$$
and then the third condition $\wt X_3\Psi=0$ is written
$$
v_1\pd{}{u_1}+v_2\pd{}{u_2}-u_1\pd{}{v_1}-u_2\pd{}{v_2}\,,
$$
for which the characteristic system is
$$
\frac{du_1}{v_1}=\frac{du_2}{v_2}=\frac{dv_1}{-u_1}=\frac{dv_2}{-u_2}
$$
from where we find the first integrals
$$
u_1^2+v_1^2=(x_0-x_1)^2+(y_0-y_1)^2=C_1\,,\qquad u_2^2+v_2^2=(x_0-x_2)^2+(y_0-y_2)^2=C_2
$$
which determine the superposition foliation and provide us with a
superposition rule for the given system involving only two
particular solutions (i.e. with $m=2$).}
\end{ex}

\section{Lie systems in Lie groups and homogeneous spaces}

Let us consider now the particular case $m=1$, i.e. when a single
solution is enough to obtain any other solution. Let us assume
additionally that $\mathcal{F}=\mathcal{F}_0$, i.e. that the
superposition rule is unique as a foliation. This means that $r=n$
and the vector fields $X_1,\dots,X_n$ generically span $\sT N$.
Assume for simplicity that they span $\sT N$ globally and are
complete vector fields. Since these vector fields close on an
$n$-dimensional Lie algebra $\goth{g}$, we conclude that there is a
transitive action on $N$ of the simple-connected $n$-dimensional
Lie group $G$ associated with $\goth{g}$, so that $N=G/H$ with a
discrete subgroup $H$ of $G$ and the foliation
$\mathcal{F}=\mathcal{F}_0$ is generated by the fundamental vector
fields of the $G$-action. If $H$ is trivial and we consider the
standard action of $G$ on itself by left translations $L_g$, the
vector fields $X_i$ are just right-invariant vector fields on $G$.
As a superposition function corresponding to
$\mathcal{F}=\mathcal{F}_0$ we can choose the group multiplication
$\Phi:G\times G\to G$,
$\zF(g_{(1)},k)=g_{(1)}k$. In this case
$\Psi(g_{(0)},g_{(1)})=g_{(1)}^{-1}\,g_{(0)}$ is left invariant
$\Psi(g'g_{(0)},g'g_{(1)})=\Psi(g_{(0)},g_{(1)})$.

Conversely, given a Lie system defined by  a $t$-dependent vector
field of the form $$Y(t,g)=\sum_{\alpha=1}^r b_\alpha(t)\,
X^\ri_\alpha(g)\,,$$ where $X^\ri_\za$ is a basis of
right-invariant vector fields, then the projectability condition
is satisfied and there is a uniquely defined superposition rule.
Note however that if the vector fields $X^\ri_\za$ generate a
smaller Lie subalgebra, the superposition rule is not unique.

Let $\{a_1,\ldots,a_n\}$ be a basis in $T_eG$. This linear space
can be identified with the Lie algebra $\goth{g}$ of $G$, the set
of left invariant vector fields on $G$: for each $a\in T_eG$ let
$X_a^\li$ denote the corresponding left-invariant vector field in
$G$, given by $X_{a}^\li(g)=L_{g*e}a$. Similarly, $X_a^\ri$
denotes the right-invariant vector field in $G$ given by
$X_{a}^\ri(g)=R_{g*e}a$. A curve in $ T_eG$,
$$
a(t)=\sum_{\alpha=1}^n b_\alpha(t)\, a_\alpha\ ,$$  gives rise to
a $t$-dependent vector field on $G$
\be
X^\ri(t,g)=X^\ri_{a(t)}(g)=\sum_{\alpha=1}^nb_\alpha(t) \,
X^\ri_{\alpha}(g)\ ,\label{engrupo}
\ee
where in the right hand side $X_ \alpha^\ri$ is a shorthand
notation for $X^\ri_{a_\alpha}$. The associated system of
differential equations determining the integral curves of such a
$t$-dependent vector field  reads
\be
\dot g(t)=\sum_{\alpha=1}^n\, b_\alpha(t) \, X^\ri_{\alpha}(g(t))\
,\label{eqgroup}
\ee
and applying $R_{g^{-1}(x)*g(x)}$ to both sides of (\ref{eqgroup})
we  find the equation
\begin{equation}
R_{g^{-1}(t)*g(t)}\dot g(t)= \sum_{\alpha=1}^nb_\alpha(t)a_\alpha=
a(t)\ ,\label{eqfund}
\end{equation}
that with some abuse of notation we will write
\begin{equation}
(\dot g\, g^{-1})(t)=a(t)\ .
\end{equation}
The solution of this equation  starting from the neutral element
can be solved by making use of a generalization of the method
developed by Wei and Norman \cite{WN1,{WN2}} (see e.g. \cite{CarMarNas}
and \cite{CGM01}) for solving an analogous  linear problem. A
simpler situation is when $a(t)$ takes values not in the full Lie
algebra ${\goth{g}}=T_eG$ but in a subalgebra.

Now, let $H$ be an arbitrary closed subgroup of $G$ and consider
the homogeneous space $N=G/H$. Then, $G$ can be seen as a
principal bundle $\tau: G\ra G/H$. Moreover, it is also known that
the right-invariant vector fields $X^\ri_\alpha$ are
$\tau$-projectable and the $\tau$-related vector fields in $N$ are
the fundamental vector fields $-X_\alpha=- X_{a_\alpha}$
corresponding to the natural left action of $G$ on $N$,
$\tau_{*g}X_\alpha^\ri(g)=-X_\alpha(gH)$. In this way we can
associate with the Lie system on the group $X$ given by
(\ref{engrupo}) a Lie system on $N$:
\begin{equation}
\bar X(t,x)=-\sum_{\alpha=1}^nb_\alpha(t)\,X_\alpha(x)\
.\label{LSsystem}
\end{equation}
Therefore, the integral curve  of (\ref{LSsystem}) starting from
$x_0$ are given by $x(t)=\Phi(g(t),x_0)$, with $g(t)$ being the
solution of (\ref{eqfund}) with $g(0)=e$.

Let us note that even if the original $t$-dependent vector field
on $N$ is projectable to $N_0$ by means of a submersion $\zp:N\ra
N_0$, the superposition rule cannot be projected in general and
the number $m$ of solutions appearing in the superposition rule
changes. For instance, in the case of Riccati equation a
fundamental set is made of three solutions, while for the linear
realization of $SL(2,\mathbb{R})$ on $\mathbb{R}^2$ only two
solutions are needed, and for $SL(2,\mathbb{R})$ acting on itself
only one solution is sufficient.

In the particular case of a Lie system $\dot g\, g^{-1}=a$ in a
Lie group $G$, it was recently shown \cite{CarRamGra} that the
knowledge of a particular solution of the corresponding system in
a homogeneous space for $G$ reduces the problem to one on the
isotopy group of a point in the homogeneous space. So, if $x(t)$
is the  particular solution of the associated Lie system in a
homogeneous space starting from $x_0$, then we can choose a curve
$\bar g(t)$ such that $\Phi(\bar g(t),x_0)=x(t)$ and there should
exist a curve $h(t)\in G_{x_0}$ such that $g(t)=\bar g(t)\,h(t)$.
Such curve $h(t)$ is a solution of the Lie equation in $G_{x_0}$,
$\dot h \, h^{-1}={\rm Ad\,}\bar g^{-1}(a+\dot{\bar g}\bar
g^{-1})$. Therefore, finding a solution of such equation in the
subgroup $G_{x_0}$, we can recover the solution $g(t)$ of the Lie
system in $G$ as $g(t)=\bar g(t)\, h(t)$. Using a new solution
starting from a new point, the problem is further reduced and
therefore with a number of known solutions we can directly write
the general solution.

Another relevant case is when there exists an equivariant map
$F:N_1\to N_2$ between two homogeneous spaces of a Lie group $G$.
In this case the corresponding fundamental fields are $F$-related
and then the image under $F$ of an integral curve of a Lie system
in $M_1$ is an integral curve of the corresponding system in
$M_2$. A very simple example is  the following: the function
$F:{\Bbb R}^2-\{(0,0)\}\to{\overline{\Bbb R}}={\Bbb
R}\cup\{\infty\}$ given by
$$F(x_1,x_2)=\left\{
\begin{array}{rl}\dfrac{x_1}{x_2}&
{\rm if}\quad x_2\ne 0 \\
\infty &{\rm if}\quad x_2= 0\end{array}\right.
$$
is equivariant with respect to the linear action of the Lie  group
$SL(2,{\R})$ on ${\Bbb R}^2-\{(0,0)\}$, and its  action on the
completed real line $\overline{\R}$, given by
\begin{eqnarray}
\Phi(A,x)&=&{\frac{\alpha\, x+\beta}{\gamma\, x+\delta}},\ \ \
\mbox{if}\ x\neq-{\frac{\delta}{\gamma}},
\nonumber\\
\Phi(A,\infty)&=&{\frac{\alpha}{\gamma}}\ ,\ \ \ \
\Phi(A,-{{\delta}/{ \gamma}})=\infty, \nonumber\end{eqnarray} when
$A$ is the matrix given by
\begin{equation}
A=\matriz{cc}{{\alpha}& {\beta}\\{\gamma}& {\delta}} \,\in
SL(2,{\R})\ .
\end{equation}
Consequently, given an integral curve of the system of
differential equations
$$
\left\{
\begin{array}{rcl}
\dot x_1&=&\dfrac 12\,b_2\,x_1+b_1\,x_2\,,\\&&\\
\dot x_2&=&-b_3\,x_1-\dfrac 12b_2\,x_2\,,
\end{array}\right.
$$
the curve $x(t)=x_1(t)/x_2(t)$ is an integral curve of the Riccati
equation
$$\dot x=b_1+b_2\, x+b_3\, x^2\ .
$$
This is precisely the method by which Riccati arrived to this last
equation.

\section{Partial superposition rules}

Finding new solutions from know ones is a very usual method in the theory of
both ordinary and partial differential equations and this procedure
has many
applications in physics. In order to deal in a geometric way with this problem
we introduce next a concept generalizing those of nonlinear
superposition introduced so far and  the one in \cite{AI70} as well as that 
of connecting function
used in \cite{JA67} (see also \cite{SL70} for some examples).

A {\it partial superposition rule} of rank $s$ of  $m$ solutions
for the system  of ordinary differential equations (\ref{nonasys})
is given by a  function
$\Phi:{\R}^{n\,m+s}\to {\R}^n$,
\be
x=\Phi(x_{(1)}, \ldots,x_{(m)};k_1,\ldots,k_s)\ ,\label{superpfs}
\ee
such that if  $\{x_{(a)}(t)\mid a=1,\ldots,m\}$ is a set of $m$
particular solutions of the system (\ref{nonasys}), then, at least for
sufficiently small $t$,
\be
x(t)=\Phi(x_{(1)}(t), \ldots,x_{(m)}(t);k_1,\ldots,k_s)\ ,\label{superpfts}
\ee
is also a solution of the system (\ref{nonasys}),
where
$k=(k_1,\ldots,k_s)$ is a set of $s$  arbitrary constants.

Note that this new concept
reduces to the previously considered superposition rule for $s=n$
and coincides with a $s$-parameter family of connecting functions in the sense
of \cite{JA67}.

Given such a partial superposition function, there is a non-uniquely defined
 function $\Psi:{\R}^{n(m+1)}\to {\R}^n$ such that
$$\Psi^i(x_{(0)},x_{(1)}, \ldots,x_{(m)})=\left\{\begin{array}{rl}k_i&{\rm if}\quad
    i\leq s\\0&{\rm if}\quad
    i> s\end{array}\right.
$$

The last $(n-s)$ equations are restrictions defining a submanifold $M$ of
codimension $(n-s)$ of ${\R}^{n(m+1)}$, i.e. of dimension $m\,n+s$, and the other equations define a
foliation of codimension $s$ in $M$. Now, following the same procedure as
in Section 2 we will arrive to a distribution in ${\R}^{n(m+1)}$
spanned by the $t$-dependent vector fields $Y_a$ defined by $(\ref{Ya})$ which
provide us with a distribution in $M$, because they are tangent to $M$,
whose
integral leaves are $n\, m$-dimensional and each
leaf is fixed by the choice of $s$ constants $k_1,\ldots,k_s$.

Moreover, the restriction ${\rm pr}_{|M}$
of ${\rm pr}$ on the submanifold $M$ defines a
subbundle of  ${\rm pr}:{\R}^{n(m+1)}\to {\R}^{n\,m}$ and
establishes
diffeomorphisms among the different leaves and
allows us to identify among them
 the leaves of the foliation defined by the prolongation of the given
non-autonomous system.

Conversely, if $M$ is a submanifold of ${\R}^{n(m+1)}$ of codimension
 $(n-s)$ which defines a subbundle of ${\rm pr}:{\R}^{n(m+1)}\to {\R}^{n\,m}$
such that the distribution defined in ${\R}^{n(m+1)}$ by the
 prolongation of vector fields is also a distribution in $M$, i.e. such vectors
 are tangent to $M$, and the restriction ${\rm pr}_{|M}$ provides us with
diffeomorphisms among the different leaves
allowing  us to identify among them
 the leaves of the foliation defined by the prolongation of the given
non-autonomous system. Such diffeomorphisms can be used to define a
superposition rule of $m$ solutions involving $s$ constants,

If for instance we consider the linear system
\begin{eqnarray}\frac{dx^1}{dt}&=&a_{11}(t)\,x^1+a_{12}(t)\,x^2\\
\frac{dx^2}{dt}&=&a_{21}(t)\,x^1+a_{22}(t)\,x^2
\end{eqnarray}
then  it admits a superposition function of rank one and involving one
particular solution, $F(x_{(1)};k)=k\ x_{(1)}$, which determines the
three-dimensional subbundle $M$ of ${\rm pr}:\mathbb{R}^4\to \mathbb{R}^2$
defined by the restriction to the subset given by the relation
$$x^1\ x_{(1)}^2-x^2\ x_{(1)}^1=0\ ,
$$
which is endowed with a foliation: each leaf is characterized by a real number
$k$ and is defined on the set of points $(x^1,x^2,x_{(1)}^1,x_{(1)}^2)$ such
that  $x^1\ x_{(1)}^2-x^2\ x_{(1)}^1=0$.

However we have also a superposition function of rank one but involving two
constants:
$$F(x_{(1)}, x_{(2)};k)=x_{(1)}+k\ x_{(2)}\ ,
$$

The subbundle now will be defined by
$$x_{(2)}^1(x^2-x_{(1)}^2)-x_{(2)}^2(x^1-x_{(1)}^1)=0\ .
$$

\section{Superposition rules for PDE's}

Consider now the system of first-order PDE's of the form:
\begin{equation}
\pd{x^i}{t^a}=Y^i_a(t,x)\,,\qquad  x\in{\R}^n,\
t=(t^1,\ldots,t^s)\in {\R}^s\,,\label{lpdesys}
\end{equation}
whose solutions are maps $x(t):{\R}^s\to {\R}^n$.

A particular case of (\ref{lpdesys})  when $s=1$ is (\ref{nonasys}). The main
difference of (\ref{lpdesys}) with respect to  (\ref{nonasys}) is that for
$s>1$ we have no, in general, existence of a solution with a given initial
value $x(0)\in {\R}^n$. For a better understanding of this problem, let us put
(\ref{lpdesys}) in a more general and geometric framework.

For a manifold $N$ of dimension $n$ consider the trivial fibre  bundle
$$P_N^s={\R}^s\times N\to {\R}^s\,.$$
A connection $\bar Y$ in this bundle is a horizontal distribution
in $TP_N^s$. i.e. an $s$-dimensional distribution transversal to
the fibres. It is determined by horizontal lifts of the coordinate
vector fields $\partial/\partial t^a$ in ${\R}^s$ which read
$$\bar Y_a=\pd{}{t^a}+Y_a(t,x)$$
with
$$Y_a(t,x)=Y^i_a(t,x)\pd{}{x^i}\,.
$$ Thus, the solutions of (\ref{lpdesys}) can be identified  with integral
submanifolds of the distribution $\bar Y$,
$$(t,Y(t))\,,\qquad t\in {\R}^s\,.
$$
It is now clear that there is an (obviously unique) solution of
(\ref{lpdesys}) for every initial data if and only if the
distribution $\bar Y$ is integrable, i.e. the connection  has a
trivial curvature. This means that
$$[\bar Y_a,\bar Y_b]=\sum_{c=1}^r f_{ab}\, ^c \
\bar Y_c$$ for some functions $ f_{ab}\, ^c$ in $P_N^s$. But the
commutators $[\bar Y_a,\bar Y_b]$ are clearly vertical while $\bar
Y_c$ are linearly independent horizontal vector fields, so $
f_{ab}\, ^c=0$ which yields the integrability condition in the
form of the system of equations $ [\bar Y_a,\bar Y_b]=0$, i.e., in
local coordinates,
\begin{equation}
\pd{Y^i_b}{t^a}(t,x)-\pd{Y^i_a}{t^b}(t,x)
+\sum_{j=1}^n\left(Y^j_a(t,x)\pd{Y^i_b}{x^j}(t,x)-Y^j_b(t,x)\pd{Y^i_a}{x^j}(t,x)
\right)=0\,.\label{integcond}
\end{equation}
Let us assume now that we work with a system of  first-order PDE's
of the form (\ref{lpdesys}) and satisfying the integrability
conditions (\ref{integcond}). Then we are sure that, for a given
initial value, there is a unique solution of (\ref{lpdesys}). Now,
we can think about superposition rules for such solutions. It is,
however, completely obvious that  the concepts of superposition
rules we have developed can be applied with no real changes to our
case of PDE's. In the formula  (\ref{superpf}) we should now think
that $t$ is not a real parameter but $t\in{\R}^s$. The only
difference when passing to the foliation induced by the
superposition function $\Psi$ is that we differentiate
(\ref{defPsi}) not with respect to the simple parameter $t$ but
with respect to all $t^a$. Therefore, the proposition 1 takes the
form:

\begin{prop}
Giving a superposition rule for the system (\ref{lpdesys}) satisfying the
integrability condition (\ref{integcond}) is equivalent with giving a connection in the
bundle ${\rm pr}:N^{(m+1)}\to N^m$
with a zero curvature and for which the diagonal prolongations $\widetilde
Y_a(t)$ of all the vector fields $Y_a(t)$, $t\in {\R}^s$, $a=1,\ldots, s$, are
horizontal.
\end{prop}
Also the proof of Lie theorem remains unchanged. Therefore we get
the following analog of the Lie theorem for PDE's:
\begin{theo}
The system (\ref{lpdesys}) of PDE's defined on a manifold $N$ and
satisfying the integrability condition (\ref{integcond}) admits a
superposition rule if and only if the vector fields $Y_a(t,x)$ on
$N$ depending on the parameter $t\in{\R}^s$, can be written
locally in the form
\be\label{yy}Y_a(t,x)=\sum_{\alpha=1}^ru_a^\alpha(t)X_\alpha(x)\,,\qquad
a=1,\ldots s\,,
\ee
where the vector fields $X_\alpha$ close on a finite-dimensional real Lie
algebra,
i.e. there exist $r^3$ real constants $c_{\alpha\beta}\,^\gamma$ such that
$$
[X_\alpha,X_\beta]=\sum_{\gamma=1}^rc_{\alpha\beta}\,^\gamma\,X_\gamma\,.
$$
\end{theo}
Let us observe that the integrability condition for $Y_a(t,x)$ of
the form (\ref{yy}) can be written as
$$\sum_{\za,\zb,\zg=1}^r\left[(u^\zg_b)'(t)-(u^\zg_a)'(t)+
u^\za_a(t)u^\zb_b(t)c^\zg_{\za\zb}\right]X_\zg=0.$$

\begin{ex}{\rm
Consider the following system of partial differential equations on
$\R^2$ associated with the $sl(2,\R)$-action on $\R$ represented
by vector fields (\ref{ric}):
\beas u_x&=&a(x,y)u^2+b(x,y)u+c(x,y)\,,\\
u_y&=&d(x,y)u^2+e(x,y)u+f(x,y)\,.
\eeas
This equation can be written in the form of a `total differential
equation'
$$(a(x,y)u^2+b(x,y)u+c(x,y))\xd x+(d(x,y)u^2+e(x,y)u+f(x,y))\xd
y=\xd u\,.$$ The integrability condition just says that the
one-form
$$\zw=(a(x,y)u^2+b(x,y)u+c(x,y))\xd x+(d(x,y)u^2+e(x,y)u+f(x,y))\xd
y$$ is closed for arbitrary function $u=u(x,y)$. If this is the
case, then there is a unique solution with the initial condition
$u(x_0,y_0)=u_0$ and there is a superposition rule giving a
general solution as a function of three independent solutions
exactly as in the case of Riccati equation:
$$\frac{(u-u_{(1)})(u_{(2)}-u_{(3)})}{(u-u_{(2)})(u_{(1)}-u_{(3)})}=k\,,$$
or
$$u=\frac{(u_{(1)}-u_{(3)})u_{(2)}k+u_{(1)}(u_{(3)}-u_{(2)})}
{(u_{(1)}-u_{(3)})k+(u_{(3)}-u_{(2)})}\,.
$$
}\end{ex}

\section{Concluding remarks}

In this paper we have identified  and solved a gap present in previous
proofs of the necessary and
sufficient conditions for the existence of a superposition rule for 
ordinary differential equations.
In doing this we  provided the superposition rule with a much better
geometrical interpretation which
allows us  to exhibit many interesting  properties. For instance, it is
now clear by inspection that
the vector  field as given in the Theorem 1  may be multiplied by a
function of time, i.e. we may perform
a re-parametrization in time, and still get an equation admitting a
superposition principle. In this way we
find that if  the superposition rule holds true for  a vector field
field, it is also true for all re-parametrized
vector fields. In particular, it would allow us  to write a superposition rule
also for vector fields which may be
reduced to autonomous ones via re-parametrization.
The new  geometrical interpretation also paves the way to a proper treatment of the
superposition rule for partial differential equations.
We hope to be able to extend the treatment to field theory and perhaps
be able to get interacting
field theories out of free ones very much as it happens for ordinary
differential equations. Indeed it is known
that Riccati equation obtains from a linear system. In previous papers it has been
shown how to cast completely integrable systems in a generalized version of Lie-Scheffers
systems, as most completely integrable systems do arise as reduction of simple systems
we hope  to be able
to show in general
that systems which allow for a superposition rule may be derived from
`simple ones', both for ordinary  and partial differential equations.

\section{Acknowledgment.}
This work was partially supported by the INFN-MEC
collaboration agreement no 06-23, 
the  research projects BFM2003-02532,
DGA-GRUPOS CONSOLIDADOS E24/1 and 
PRIN SINTESI.

\end{document}